\documentclass[12pt,fleqn]{article}
\pdfoutput=1
\usepackage[dvips]{epsfig}
\usepackage{amsmath,amssymb,amsfonts}
\usepackage{tikz}
\usepackage{tikz,tikz-cd}
\usepackage{hyperref} 

\numberwithin{equation}{subsection}
\numberwithin{figure}{subsection}

\newtheorem{theorem}{Theorem}[subsection]

\newtheorem{remark}[theorem]{Remark}

\newtheorem{examples}[theorem]{Examples}

\def\pr{\operatorname{pr}}%
\def\spec{\operatorname{Spec}}%

\newcommand{\qed}{\nobreak \ifvmode \relax \else
      \ifdim\lastskip<1.5em \hskip-\lastskip
      \hskip1.5em plus0em minus0.5em \fi \nobreak
      \vrule height0.75em width0.5em depth0.25em\fi}

 0 3

\begin{document}

\title{On an Internal Dependence
\\ of Simultaneous Measurements}

\author{Valentin Vankov Iliev\\
Institute of Mathematics and Informatics\\
Bulgarian Academy of Sciences\\
Sofia, Bulgaria\\
e-mail: viliev@math.bas.bg\\
 }

\maketitle

\begin{abstract} In this paper we show that
there exists an internal dependence of the simultaneous
measurements made by the two pairs of linear polarizers operated
in each leg of the apparatus in Aspect's version of
Einstein-Podolsky-Rosen-Bohm \emph{Gedankenexperiment}. The
corresponding Shannon-Kolmogorov's information flow linking a
polarizer from one leg to a polarizer from the other leg is
proportional to the absolute value of this function of dependence.
It turns out that if Bell's inequality is violated, then this
information flow is strictly positive, that is, the experiment
performed at one leg is informationally dependent on the
experiment at the other leg. By throwing out the sign of absolute
value, we define the signed information flow linking a polarizer
from one leg to a polarizer from the other leg which, in turn,
reproduces the probabilities of the four outcomes of the
simultaneous measurements, predicted by quantum mechanics. We make
an attempt to illustrate the seeming random relation between the
total information flow, the total signed information flow, and the
violation of Bell's inequality in terms of a kind of uncertainty
principle.

\end{abstract}

\section{Introduction, Notation}

\label{1}

\subsection{Introduction}

\label{1.1}

In the context of the bipartite quantum system that describes
Aspect's optical version of Einstein-Podolsky-Rosen-Bohm
\emph{Gedankenexperiment} (see ~\cite{[1]} and~\cite{[20]}), we
consider the pairs of linear polarizers operated in each leg of
the apparatus as pairs of self-adjoined linear operators
\[
A_{\mu_i}=\left(
\begin{array}{ccccc}
\cos\mu_i &\sin\mu_i \\
\sin\mu_i & -\cos\mu_i \\
 \end{array}
\right), B_{\nu_j}=A_{\nu_j},
\]
where $\mu_i, \nu_j\in [0,\pi]$, $i,j=1,2$, are the angles of the
polarizers. Note that each pair has a time switch which
interchanges polarizers, the corresponding time being shorter than
the time necessary for a light signal to travel from one of the
pairs of polarizers to the other (Einstein locality assumption for
independence).

Each pair of operators $A_{\mu_1}, A_{\mu_2}$ and $B_{\nu_1},
B_{\nu_2}$ acts on the state space of the corresponding quantum
subsystem (a unitary plane). By tensoring with the unit operator
on the other plane, we obtain two pairs of self-adjoined linear
operators $\mathcal{A}_{\mu_1}$, $\mathcal{A}_{\mu_2}$ and
$\mathcal{B}_{\nu_1}$, $\mathcal{B}_{\nu_2}$ with spectre
$\{1,-1\}$ on the state space of the whole quantum system (tensor
product of the two unitary planes). Moreover, for each $i,j=1,2,$
the operators $\mathcal{A}_{\mu_i}$ and $\mathcal{B}_{\nu_j}$
commute because the state space of the whole quantum system has an
orthonormal frame consisting of eigenvectors of both operators. In
this case the corresponding measurements are said to be
\emph{simultaneous}.

In accord with the axiom of quantum mechanics about the
observables, after fixing initial state we can consider the
members of this frame as outcomes of a sample space with
probability assignment consisting of probabilities predicted by
this axiom. Moreover, with an abuse of the language, we can also
consider the operators as random variables with range $\{1,-1\}$
on this sample space. Under the condition that the singlet state
is initial, any one of these random variables has probability
distribution $(\frac{1}{2},\frac{1}{2})$. Moreover, if
$\mu\in\{\mu_1,\mu_2\}$ and  $\nu\in\{\nu_1,\nu_2\}$, then
\[
\pr((\mathcal{A}_\mu=1)\cap (\mathcal{B}_\nu=1))=
\pr((\mathcal{A}_\mu=-1)\cap
(\mathcal{B}_\nu=-1))=\frac{1}{2}\sin^2\left(\frac{\mu-\nu}{2}\right),
\]
\[
\pr((\mathcal{A}_\mu=1)\cap (\mathcal{B}_\nu=-1))=
\pr((\mathcal{A}_\mu=-1)\cap (\mathcal{B}_\nu=1))=
\frac{1}{2}\cos^2\left(\frac{\mu-\nu}{2}\right).
\]
Therefore, the product of random variables
$\mathcal{A}_\mu\mathcal{B}_\nu$ has probability distribution
$\left(\sin^2\left(\frac{\mu-\nu}{2}\right),\cos^2\left(\frac{\mu-\nu}{2}\right)\right)$
and expected value
$\mathcal{E}(\mathcal{A}_\mu\mathcal{B}_\nu)=-\cos(\mu-\nu)$.

On the other hand, the joint experiment (see~\cite[Part I, Section
6]{[35]}) of the binary trials
$\mathfrak{A}_\mu=(\mathcal{A}_\mu=1)\cup(\mathcal{A}_\mu=-1)$ and
$\mathfrak{B}_\nu=(\mathcal{B}_\nu=1)\cup(\mathcal{B}_\nu=-1)$
produces the probability distribution
\[
\left(\frac{1}{2}\sin^2\left(\frac{\mu-\nu}{2}\right),\frac{1}{2}\cos^2\left(\frac{\mu-\nu}{2}\right),
\frac{1}{2}\cos^2\left(\frac{\mu-\nu}{2}\right),\frac{1}{2}\sin^2\left(\frac{\mu-\nu}{2}\right)\right)
\]
with Boltzmann-Shannon entropy $E(\theta_{\mu,\nu})$, where
$E(\theta)=
-2\theta\ln\theta-2(\frac{1}{2}-\theta)\ln(\frac{1}{2}-\theta)$
and
$\theta_{\mu,\nu}=\frac{1}{2}\sin^2\left(\frac{\mu-\nu}{2}\right)$.
We extend the function $E(\theta)$, $\theta\in (0,\frac{1}{2})$,
as continuous on the closed interval $[0,\frac{1}{2}]$.

By modifying the entropy function $E(\theta)$, we obtain the
strictly increasing \emph{degree of dependence function} $e\colon
[0,\frac{1}{2}]\to [-1,1]$, which mimics the regression
coefficient (see~\cite[5.2]{[25]}).

It turns out that the average quantity of information
$I(\mathfrak{A}_\mu,\mathfrak{B}_\nu)$ (see~\cite[\S 1]{[10]}) of
one of the experiments $\mathfrak{A}_\mu$ and $\mathfrak{B}_\nu$,
relative to the other can be found by the formula
$I(\mathfrak{A}_\mu,\mathfrak{B}_\nu)=|e(\theta_{\mu,\nu})|\ln 2$.
We can consider $I(\mathfrak{A}_\mu,\mathfrak{B}_\nu)$ as a
measure of the flow carrying information between these two binary
trials (see~\cite[5.3]{[25]}). Since $s$ is an invertible
function, the corresponding \emph{signed information flow}
$I^{\left(s\right)}(\mathcal{A}_\mu,\mathcal{B}_\nu)(\theta)=e(\theta)\ln
2$ replicates the probability distribution~\eqref{10.1.5} produced
by quantum mechanics.

In terms of Aspect's experiment, the sum
$I(\mathfrak{A},\mathfrak{B})=\sum_{i,j}^2
I(\mathfrak{A}_{\mu_i},\mathfrak{B}_{\nu_j})$ (called total
information flow) can be thought about as a measure of the flow
carrying information between the two pairs of polarizers. In his
paper~\cite{[5]} John Bell deduced under the assumptions of
"locality" and "realism" that if measurements are performed
independently (Einstein locality assumption for independence) on
the two separated particles (photons in Aspect's experiment) of an
entangled pair, then the assumption that the outcomes depend upon
"hidden variables" implies constraint condition called Bell's
inequality (see Subsection~\ref{15.1}). It comes out that if
Bell's inequality is violated, then the total information flow is
strictly positive. In other words, in this case there exists an
informational dependence between the two legs of apparatus.

In the end of the paper we discuss the relation between the
information flow $I(\mathfrak{A},\mathfrak{B})$ and the violation
of Bell's inequality. Using Examples~\ref{15.1.1} and the Java
program from the link that can be found there, we note that this
relation is subject to a kind of uncertainty principle.

\subsection{Notation}

\label{1.5}

\noindent $\mathcal{H}$: $2$-dimensional unitary space with inner
product $\langle x| y\rangle$ which is linear in the second slot
and anti-linear in the first slot;

\noindent $\mathbb{I}=\mathbb{I}_{\mathcal{H}}$: the identity
linear operator on $\mathcal{H}$;

\noindent $\mathcal{H}^{\otimes 2}=\mathcal{H}\otimes\mathcal{H}$:
the unitary tensor square with inner product $\langle x_1\otimes
x_2| y_1\otimes y_2\rangle=\langle x_1| y_1\rangle\langle x_2|
y_2\rangle$;

\noindent $\mathcal{U}^{\left(2\right)}$: the unit sphere in
$\mathcal{H}^{\otimes 2}$;

\noindent $\spec(A)$: the real spectre of a self-adjoined linear
operator $A$ on $\mathcal{H}$ with trace zero, having the form
$\spec(A)=\{\lambda_1^{\left(A\right)},\lambda_2^{\left(A\right)}\}$,
$\lambda_1^{\left(A\right)}+\lambda_2^{\left(A\right)}=0$;

\noindent $u^{\left(A\right)}=\{u_1^{\left(A\right)},
u_2^{\left(A\right)}\}$: the orthonormal frame for $\mathcal{H}$,
formed by the corresponding eigenvectors of $A$;

\noindent $\mathcal{H}_i^{\left(A\right)}$: the eigenspaces
$\mathbb{C}u_i^{\left(A\right)}$ of $A$, $i=1,2$;

\section{Self-Adjoint Operators on $\mathcal{H}$}

\label{5}

\subsection{Two Special Commuting Operators}

\label{5.10}

We fix an orthonormal frame $h=\{h_1,h_2\}$ for $\mathcal{H}$ and
identify the self-adjoined operators with their matrices with
respect to $h$. For any $\mu\in [0,\pi]$ we denote by $A_\mu$ the
self-adjoined operator
\[ \left(
\begin{array}{ccccc}
\cos\mu &\sin\mu \\
\sin\mu & -\cos\mu \\
 \end{array}
\right).
\]
We have $\lambda_1^{\left(A_\mu\right)}=1,
\lambda_2^{\left(A_\mu\right)}=-1$, and
\[
u_1^{\left(A_\mu\right)}=(\cos\frac{\mu}{2})h_1+(\sin\frac{\mu}{2})h_2,
u_2^{\left(A_\mu\right)}=(-\sin\frac{\mu}{2})h_1+(\cos\frac{\mu}{2})h_2.
\]
For any $\nu\in [0,\pi]$ we set $B_\nu=A_\nu$.

Note that $\{h_1\otimes h_1, h_1\otimes h_2, h_2\otimes h_1,
h_2\otimes h_2\}$ and $u^{\left(A_\mu\right)}\otimes
u^{\left(B_\nu\right)} =\{u_1^{\left(A_\mu\right)}\otimes
u_1^{\left(B_\nu\right)},u_1^{\left(A_\mu\right)}\otimes
u_2^{\left(B_\nu\right)},u_2^{\left(A_\mu\right)}\otimes
u_1^{\left(B_\nu\right)},u_2^{\left(A_\mu\right)}\otimes
u_2^{\left(B_\nu\right)}\}$ are orthonormal frames for
$\mathcal{H}^{\otimes 2}$.

Let us set $\mathcal{A}_\mu=A_\mu\otimes\mathbb{I}$,
$\mathcal{B}_\nu=\mathbb{I}\otimes B_\nu$. It is a straightforward
check that the last two linear operators on $\mathcal{H}^{\otimes
2}$ are also self-adjoined with
$\lambda_1^{\left(\mathcal{A}_\mu\right)}=
\lambda_1^{\left(\mathcal{B}_\nu\right)}=1$,
$\lambda_2^{\left(\mathcal{A}_\mu\right)}=
\lambda_2^{\left(\mathcal{B}_\nu\right)}=-1$, the
$\lambda_i^{\left(\mathcal{A}_\mu\right)}$-eigenspace
$\mathcal{H}_i^{\left(\mathcal{A}_\mu\right)}=
\mathcal{H}_i^{\left(A_\mu\right)}\otimes\mathcal{H}$ has
orthonormal frame $\{u_i^{\left(A_\mu\right)}\otimes
u_1^{\left(B_\nu\right)},u_i^{\left(A_\mu\right)}\otimes
u_2^{\left(B_\nu\right)}\}$, and the
$\lambda_i^{\left(\mathcal{B}_\nu\right)}$-eigenspace
$\mathcal{H}_i^{\left(\mathcal{B}_\nu\right)}=
\mathcal{H}\otimes\mathcal{H}_i^{\left(B_\nu\right)}$ has
orthonormal frame $\{u_1^{\left(A_\mu\right)}\otimes
u_i^{\left(B_\nu\right)},u_2^{\left(A_\mu\right)}\otimes
u_i^{\left(B_\nu\right)}\}$, $i=1,2$.

Since $u^{\left(A_\mu\right)}\otimes u^{\left(B_\nu\right)}$ is an
orthonornal frame of $\mathcal{H}^{\otimes 2}$ consisting of
eigenvectors of both $\mathcal{A}_\mu$ and $\mathcal{B}_\nu$, then
the last two operators commute.

Let $\psi\in\mathcal{U}^{\left(2\right)}$ and let
$S(\psi;\mathcal{A}_\mu,\mathcal{B}_\nu)$ be the sample space with
set of outcomes $u^{\left(A_\mu\right)}\otimes
u^{\left(B_\nu\right)} =\{u_1^{\left(A_\mu\right)}\otimes
u_1^{\left(B_\nu\right)},u_1^{\left(A_\mu\right)}\otimes
u_2^{\left(B_\nu\right)},u_2^{\left(A_\mu\right)}\otimes
u_1^{\left(B_\nu\right)},u_2^{\left(A_\mu\right)}\otimes
u_2^{\left(B_\nu\right)}\}$ and probability assignment $\{p_{11},
p_{12},p_{21}, p_{22} \}$ with $p_{ij}=|\langle
u_i^{\left(A_\mu\right)}\otimes
u_j^{\left(B_\nu\right)}|\psi\rangle|^2$, $i,j=1,2$. With an abuse
of the language, we consider the observable $\mathcal{A}_\mu$ as a
random variable $\mathcal{A}_\mu\colon
u^{\left(A_\mu\right)}\otimes
u^{\left(B_\nu\right)}\to\mathbb{R}$,
$\mathcal{A}_\mu(u_1^{\left(A_\mu\right)}\otimes
u_j^{\left(B_\nu\right)})=\lambda_1^{\left(\mathcal{A}_\mu\right)}$,
$\mathcal{A}_\mu(u_2^{\left(A_\mu\right)}\otimes
u_j^{\left(B_\nu\right)})=\lambda_2^{\left(\mathcal{A}_\mu\right)}$,
$j=1,2$, on the sample space
$S(\psi;\mathcal{A}_\mu,\mathcal{B}_\nu)$ with probability
distribution
$p_{\mathcal{A}_\mu}(\lambda_i^{\left(A\right)})=|\langle
u_i^{\left(A_\mu\right)}\otimes
u_1^{\left(B_\nu\right)}|\psi\rangle|^2+|\langle
u_i^{\left(A_\mu\right)}\otimes
u_2^{\left(B_\nu\right)}|\psi\rangle|^2$, $i=1,2$, and
$p_{\mathcal{A}_\mu}(\lambda)=0$ for
$\lambda\notin\spec(\mathcal{A}_\mu)$. Identifying the event
$\{u_i^{\left(A_\mu\right)}\otimes
u_1^{\left(B_\nu\right)},u_i^{\left(A_\mu\right)}\otimes
u_2^{\left(B_\nu\right)}\}$ with the "event"
$\mathcal{A}_\mu=\lambda_i^{\left(\mathcal{A}_\mu\right)}$, we
have $\pr(\mathcal{A}_\mu=
\lambda_i^{\left(\mathcal{A}_\mu\right)})=|\langle
u_i^{\left(A_\mu\right)}\otimes
u_1^{\left(B_\nu\right)}|\psi\rangle|^2+|\langle
u_i^{\left(A_\mu\right)}\otimes
u_2^{\left(B_\nu\right)}|\psi\rangle|^2$, $i=1,2$.

We also consider the observable $\mathcal{B}_\nu$ as a random
variable $\mathcal{B}_\nu\colon u^{\left(A_\mu\right)}\otimes
u^{\left(B_\nu\right)}\to\mathbb{R}$,
$\mathcal{B}_\nu(u_j^{\left(A_\mu\right)}\otimes
u_1^{\left(B_\nu\right)})=\lambda_1^{\left(\mathcal{B}_\nu\right)}$,
$\mathcal{B}_\nu(u_j^{\left(A_\mu\right)}\otimes
u_2^{\left(B_\nu\right)})=\lambda_2^{\left(\mathcal{B}_\nu\right)}$,
$j=1,2$, on the sample space
$S(\psi;\mathcal{A}_\mu,\mathcal{B}_\nu)$ with probability
distribution
$p_{\mathcal{B}_\nu}(\lambda_i^{\left(A\right)})=|\langle
u_1^{\left(A_\mu\right)}\otimes
u_i^{\left(B_\nu\right)}|\psi\rangle|^2+|\langle
u_2^{\left(A_\mu\right)}\otimes
u_i^{\left(B_\nu\right)}|\psi\rangle|^2$, $i=1,2$, and
$p_{\mathcal{B}_\nu}(\lambda)=0$ for
$\lambda\notin\spec(\mathcal{B}_\nu)$. Identifying the event
$\{u_1^{\left(A_\mu\right)}\otimes
u_i^{\left(B_\nu\right)},u_2^{\left(A_\mu\right)}\otimes
u_i^{\left(B_\nu\right)}\}$ with the "event"
$\mathcal{B}_\nu=\lambda_i^{\left(\mathcal{B}_\nu\right)}$, we
have $\pr(\mathcal{B}_\nu=
\lambda_i^{\left(\mathcal{B}_\nu\right)})=|\langle
u_1^{\left(A_\mu\right)}\otimes
u_i^{\left(B_\nu\right)}|\psi\rangle|^2+|\langle
u_2^{\left(A_\mu\right)}\otimes
u_i^{\left(B_\nu\right)}|\psi\rangle|^2$, $i=1,2$.

In particular, let us set $\psi=\frac{1}{\sqrt{2}}(h_1\otimes
h_2-h_2\otimes h_1)$. We have
\[
\pr(\mathcal{A}_\mu= \lambda_i^{\left(\mathcal{A}_\mu\right)})=
\]
\[
\frac{1}{2}|\langle u_i^{\left(A_\mu\right)}|h_1\rangle\langle
 u_1^{\left(B_\nu\right)}|
h_2\rangle-\langle u_i^{\left(A_\mu\right)}|h_2\rangle\langle
u_1^{\left(B_\nu\right)}| h_1\rangle|^2+
\]
\[
\frac{1}{2}|\langle u_i^{\left(A_\mu\right)}|h_1\rangle\langle
 u_2^{\left(B_\nu\right)}|
h_2\rangle-\langle u_i^{\left(A_\mu\right)}|h_2\rangle\langle
u_2^{\left(B_\nu\right)}| h_1\rangle|^2
\]
and
\[
\pr(\mathcal{B}_\nu= \lambda_j^{\left(\mathcal{B}_\nu\right)})=
\]
\[
\frac{1}{2}|\langle u_1^{\left(A_\mu\right)}|h_1\rangle\langle
 u_j^{\left(B_\nu\right)}|
h_2\rangle-\langle u_1^{\left(A_\mu\right)}|h_2\rangle\langle
u_j^{\left(B_\nu\right)}| h_1\rangle|^2+
\]
\[
\frac{1}{2}|\langle u_2^{\left(A_\mu\right)}|h_1\rangle\langle
 u_j^{\left(B_\nu\right)}|
h_2\rangle-\langle u_2^{\left(A_\mu\right)}|h_2\rangle\langle
u_j^{\left(B_\nu\right)}| h_1\rangle|^2.
\]
Taking into account the form of the eigenvectors of the matrices
$A_\mu$ and $B_\nu$, we obtain
\[
\pr(\mathcal{A}_\mu=
\lambda_i^{\left(\mathcal{A}_\mu\right)})=\pr(\mathcal{B}_\nu=
\lambda_j^{\left(\mathcal{B}_\nu\right)})=\frac{1}{2}, i,j=1,2.
\]
We identify the intersection
$(\mathcal{A}_\mu=\lambda_i^{\left(\mathcal{A}_\mu\right)})\cap
(\mathcal{B}_\nu=\lambda_j^{\left(\mathcal{B}_\nu\right)})$ with
the event $\{u_i^{\left(A_\mu\right)}\otimes
u_j^{\left(B_\nu\right)}\}$, $i,j=1,2$, in the sample space
$S(\psi;\mathcal{A}_\mu,\mathcal{B}_\nu)$ and obtain
\[
\pr
((\mathcal{A}_\mu=\lambda_i^{\left(\mathcal{A}_\mu\right)})\cap
(\mathcal{B}_\nu=\lambda_j^{\left(\mathcal{B}_\nu\right)}))=
\]
\[
\frac{1}{2}|\langle u_i^{\left(A_\mu\right)}|h_1\rangle\langle
u_j^{\left(B_\nu\right)}|h_2\rangle-\langle
u_i^{\left(A_\mu\right)}|h_2\rangle\langle
u_j^{\left(B_\nu\right)}|h_1\rangle|^2.
\]
In particular, we have
\[
\pr
((\mathcal{A}_\mu=\lambda_1^{\left(\mathcal{A}_\mu\right)})\cap
(\mathcal{B}_\nu=\lambda_1^{\left(\mathcal{B}_\nu\right)}))=
\frac{1}{2}\sin^2\left(\frac{\mu-\nu}{2}\right),
\]
\[
\pr
((\mathcal{A}_\mu=\lambda_1^{\left(\mathcal{A}_\mu\right)})\cap
(\mathcal{B}_\nu=\lambda_2^{\left(\mathcal{B}_\nu\right)}))=
\frac{1}{2}\cos^2\left(\frac{\mu-\nu}{2}\right),
\]
\[
\pr
((\mathcal{A}_\mu=\lambda_2^{\left(\mathcal{A}_\mu\right)})\cap
(\mathcal{B}_\nu=\lambda_1^{\left(\mathcal{B}_\nu\right)}))=
\frac{1}{2}\cos^2\left(\frac{\mu-\nu}{2}\right),
\]
\[
\pr
((\mathcal{A}_\mu=\lambda_2^{\left(\mathcal{A}_\mu\right)})\cap
(\mathcal{B}_\nu=\lambda_2^{\left(\mathcal{B}_\nu\right)}))=
\frac{1}{2}\sin^2\left(\frac{\mu-\nu}{2}\right).
\]
The random variable $\mathcal{A}_\mu\mathcal{B}_\nu$ has
probability distribution
\[
p_{\mathcal{A}_\mu\mathcal{B}_\nu}(1)=
\sin^2\left(\frac{\mu-\nu}{2}\right),
p_{\mathcal{A}_\mu\mathcal{B}_\nu}(-1)=
\cos^2\left(\frac{\mu-\nu}{2}\right),
\]
and $p_{\mathcal{A}_\mu\mathcal{B}_\nu}(\lambda)=0$ for
$\lambda\neq\pm 1$. The expected value of this random variable is
$\mathcal{E}(\mathcal{A}_\mu\mathcal{B}_\nu)=-\cos(\mu-\nu)$.

\section{Entropy and Degree of Dependence}

\label{10}

\subsection{Entropy}

\label{10.1}

Now, we combine the terminology and notation of this paper with
those of~\cite{[25]}. Let us set
$A=(\mathcal{A}_\mu=\lambda_1^{\left(\mathcal{A}_\mu\right)})$,
$B=(\mathcal{B}_\nu=\lambda_1^{\left(\mathcal{B}_\nu\right)})$,
$A^c=(\mathcal{A}_\mu=\lambda_2^{\left(\mathcal{A}_\mu\right)})$,
$B^c=(\mathcal{B}_\nu=\lambda_2^{\left(\mathcal{B}_\nu\right)})$.
$\alpha=\pr(A)= \frac{1}{2}$, $\beta=\pr(B)= \frac{1}{2}$.

The pair $(A,B)$ of events in the sample space
$S(\psi;\mathcal{A},\mathcal{B})$ produces an experiment
\begin{equation}
\mathfrak{J}=(A\cap B)\cup (A\cap B^c)\cup (A^c\cap B)\cup
(A^c\cap B^c)\label{10.1.1}
\end{equation}
(cf.~\cite[I,\S 3]{[15]}) and the probabilities of its results:
\[
\xi_1=\pr(A\cap B)
=\frac{1}{2}\sin^2\left(\frac{\mu-\nu}{2}\right),\ \xi_2=\pr(A\cap
B^c)= \frac{1}{2}\cos^2\left(\frac{\mu-\nu}{2}\right),
\]
\begin{equation}
\xi_3=\pr(A^c\cap B)=
\frac{1}{2}\cos^2\left(\frac{\mu-\nu}{2}\right),\
\xi_4=\pr(A^c\cap B^c)=
\frac{1}{2}\sin^2\left(\frac{\mu-\nu}{2}\right).\label{10.1.5}
\end{equation}
The probability distribution $(\xi_1,\xi_2,\xi_3,\xi_4)$ satisfies
the linear system~\cite[4.1. (3)]{[25]} whose solutions form a
straight line with parametric representation $\xi_1=\theta$,
$\xi_2=\frac{1}{2}-\theta$, $\xi_3=\frac{1}{2}-\theta$,
$\xi_4=\theta$ in the hyperplane $\xi_1+\xi_2+\xi_3+\xi_4=1$. Note
that the parameter $\theta=\xi_1$ runs within the closed interval
$[0,\frac{1}{2}]$. The entropy of $(\xi_1,\xi_2,\xi_3,\xi_4)$ is
$E(\theta)=-\sum_{k=1}^4\xi_k(\theta)\ln(\xi_k(\theta))=
-2\theta\ln\theta-2(\frac{1}{2}-\theta)\ln(\frac{1}{2}-\theta)$
and the function $E(\theta)$ can be extended as continuous on the
interval $[0,\frac{1}{2}]$. It strictly increases on the interval
$[0,\frac{1}{4}]$, strictly decreases on the interval
$[\frac{1}{4},\frac{1}{2}]$ and has a global maximum at
$\theta=\frac{1}{4}$. In particular,
$\max_{\theta\in\left[0,\frac{1}{2}\right]}E(\theta)=E(\frac{1}{4})=2\ln
2$. Since
$\min_{\theta\in\left[0,\frac{1}{4}\right]}E(\theta)=E(0)=\ln
2=E(\frac{1}{2})=
\min_{\theta\in\left[\frac{1}{4},\frac{1}{2}\right]}E(\theta)$, we
obtain $\min_{\theta\in\left[0,\frac{1}{2}\right]}E(\theta)=\ln
2$.

\subsection{Degree of Dependence}

\label{10.5}

It is more useful to modify the entropy function, thus obtaining
the strictly increasing \emph{degree of dependence function}
$e\colon [0,\frac{1}{2}]\to [-1,1]$,
\[
e(\theta)=\left\{
\begin{array}{ccccccccccccccccccccccccccccccc}
-\frac{E(\frac{1}{4})-E(\theta)}
{E(\frac{1}{4})-E(0)} & \hbox{\rm\ if\ } 0\leq\theta\leq\frac{1}{4} \\
\frac{E(\frac{1}{4})-E(\theta)}
{E(\frac{1}{4})-E(\frac{1}{2})} & \hbox{\rm\ if\ } \frac{1}{4}\leq\theta\leq \frac{1}{2}. \\
 \end{array}
\right.
\]
Taking into account the values of extrema of entropy function, we
obtain
\[
e(\theta)=\left\{
\begin{array}{ccccccccccccccccccccccccccccccc}
-2+\frac{E(\theta)}{\ln 2}& \hbox{\rm\ if\ } 0\leq\theta\leq\frac{1}{4} \\
2-\frac{E(\theta)}{\ln 2} & \hbox{\rm\ if\ } \frac{1}{4}\leq\theta\leq \frac{1}{2}. \\
 \end{array}
\right.
\]
The events $A$ and $B$ are \emph{independent} exactly when the
entropy is maximal (equal to $2\ln 2$), that is, when
$e(\theta)=0$ and this, in turn, is equivalent to the equality
$|\mu-\nu|=\frac{\pi}{2}$. We have $e(\theta)=-1$ or $e(\theta)=1$
if and only if $|\mu-\nu|=0$ or $|\mu-\nu|=\pi$, respectively, and
in these two cases the entropy is minimal and equal to $\ln 2$.
Now, let, in addition, assume that $A$ and $B$ are events in a
sample space with equally likely outcomes. If $e(\theta)=-1$, then
one of $A$ and $B$ is a subset of the complement of the other
(maximal \emph{negative dependence}), and if $e(\theta)=1$ one of
them is a subset of the other (maximal \emph{positive
dependence}).

\subsection{The Information Flow}

\label{10.10}

The experiment $\mathfrak{J}$ from~\eqref{10.1.1} is the joint
experiment (see~\cite[Part I, Section 6]{[35]}) of two simple
binary trials: $\mathfrak{A}_\mu=A\cup A^c$ and
$\mathfrak{B}_\nu=B\cup B^c$ with $\pr(A)=\pr(B)=\frac{1}{2}$. The
\emph{average quantity of information of one of the experiments
$\mathfrak{A}_\mu$ and $\mathfrak{B}_\nu$, relative to the other},
(see~\cite[\S 1]{[10]}), is defined in this particular case by the
formula $I(\mathfrak{A}_\mu,\mathfrak{B}_\nu)(\theta)=
\xi_1(\theta)\ln 4\xi_1(\theta)+\xi_2(\theta)\ln
4\xi_2(\theta)+\xi_3(\theta)\ln 4\xi_3(\theta)+\xi_4(\theta)\ln
4\xi_4(\theta)$. The above notation is correct since the
interchanges of $A$ and $A^c$ or $B$ and $B^c$ causes permutations
of $\xi_i$'s. Thus, we obtain
$I(\mathfrak{A}_\mu,\mathfrak{B}_\nu)(\theta)=
\max_{\theta\in\left[0,\frac{1}{2}\right]}E(\theta)-E(\theta)$.
Now, the definition of the degree function $e(\theta)$ yields
immediately
$I(\mathfrak{A}_\mu,\mathfrak{B}_\nu)(\theta)=|e(\theta)|\ln 2$
for $\theta\in [0,\frac{1}{2}]$.

Translating into the language of information theory, we have
$e(\theta)=-1$ or $e(\theta)=1$ if and only if
$I(\mathfrak{A}_\mu,\mathfrak{B}_\nu)(\theta)=\max_{0\leq\tau\leq
\frac{1}{2}}I(\mathfrak{A}_\mu,\mathfrak{B}_\nu)(\tau)=\ln 2$.
Finally, we have $e(\theta)=0$ if and only if
$I(\mathfrak{A}_\mu,\mathfrak{B}_\nu)(\theta)=0$,  and under this
condition the experiments $\mathfrak{A}_\mu$ and
$\mathfrak{B}_\nu$ are said to be \emph{informationally
independent}.

\subsection{The Signed Information Flow}

\label{10.15}

Let us set
$I^{\left(s\right)}(\mathcal{A}_\mu,\mathcal{B}_\nu)(\theta)=e(\theta)\ln
2$ for $\theta\in [0,\frac{1}{2}]$ and call this quantity
\emph{average quantity of signed information of one of the events
$\mathcal{A}_\mu=\lambda_1^{\left(\mathcal{A}_\mu\right)}$ and
$\mathcal{B}_\nu=\lambda_1^{\left(\mathcal{B}_\nu\right)}$,
relative to the other}. Then
$I(\mathfrak{A}_\mu,\mathfrak{B}_\nu)=
|I^{\left(s\right)}(\mathcal{A}_\mu,\mathcal{B}_\nu)|$ and since
the function $e$ is invertible, we obtain
$\theta=e^{-1}(\frac{1}{\ln
2}I^{\left(s\right)}(\mathcal{A}_\mu,\mathcal{B}_\nu))$. In
particular, the value of the signed information flow
$I^{\left(s\right)}(\mathcal{A}_\mu,\mathcal{B}_\nu)$ reproduces
the probability distribution~\eqref{10.1.5} predicted by quantum
theory.

\section{Four Operators and Bell's Map}

\label{15}

For any $\mu_1, \mu_2, \nu_1,\nu_2\in [0,\pi]$ we consider the
self-adjoined operators $A_{\mu_i}$, $B_{\nu_j}$, $i,j=1,2$, see
Subsection~\ref{5.10}. We extend notation introduced in
Sections~\ref{5} and~\ref{10} in a natural way:
$\theta_{ij}=\frac{1}{2}\sin^2\left(\frac{\mu_i-\nu_j}{2}\right)$,
$\theta_{ij}\in [0,\frac{1}{2}]$, $\mathfrak{A}_{\mu_i}$,
$\mathfrak{B}_{\nu_j}$,
$I(\mathfrak{A}_{\mu_i},\mathfrak{B}_{\nu_j})=|e(\theta_{ij})|\ln
2$, $i,j=1,2$. The sum $I(\mathfrak{A},\mathfrak{B})=
\sum_{i,j=1}^2I(\mathfrak{A}_{\mu_i},\mathfrak{B}_{\nu_j})$ is
said to be the \emph{average quantity of information of one of the
pairs of experiments $\mathfrak{A}=\{\mathfrak{A}_{\mu_1},
\mathfrak{A}_{\mu_2}\}$ and $\mathfrak{B}=\{\mathfrak{B}_{\nu_1},
\mathfrak{B}_{\nu_2}\}$ relative to the other}, or, \emph{total
information flow}. The sum
$I^{\left(s\right)}(\mathfrak{A},\mathfrak{B})=
\sum_{i,j=1}^2I^{\left(s\right)}(\mathcal{A}_\mu,\mathcal{B}_\nu)$
is said to be the \emph{average quantity of signed information of
one of the pairs of experiments
$\mathfrak{A}=\{\mathfrak{A}_{\mu_1}, \mathfrak{A}_{\mu_2}\}$ and
$\mathfrak{B}=\{\mathfrak{B}_{\nu_1}, \mathfrak{B}_{\nu_2}\}$
relative to the other}, or, \emph{total signed information flow}.

Thus, we obtain the functions
\[
I(\mathfrak{A},\mathfrak{B})\colon [0,\pi]^4\to\mathbb{R}, (\mu_1,
\mu_2, \nu_1,\nu_2)\mapsto (\ln 2)\sum_{i,j=1}^2|e(\theta_{ij})|,
\]
and
\[
I^{\left(s\right)}(\mathfrak{A},\mathfrak{B})\colon
[0,\pi]^4\to\mathbb{R}, (\mu_1, \mu_2, \nu_1,\nu_2)\mapsto (\ln
2)\sum_{i,j=1}^2e(\theta_{ij}),
\]
which represents the intensity of information flow (respectively,
signed information flow) between the pairs of experiments
$\mathfrak{A}$ and $\mathfrak{B}$. We note that $0\leq
I(\mathfrak{A},\mathfrak{B})(\mu_1, \mu_2, \nu_1,\nu_2)\leq 4\ln
2$ and $-4\ln 2\leq
I^{\left(s\right)}(\mathfrak{A},\mathfrak{B})(\mu_1, \mu_2,
\nu_1,\nu_2)\leq 4\ln 2$.

In case $\mu_1=\mu_2=\nu_1=\nu_2$ we have
$\theta_{11}=\theta_{12}=\theta_{21}=\theta_{22}=0$,
$e(\theta_{11})=e(\theta_{12})=e(\theta_{21})=e(\theta_{22})=-1$,
hence $I(\mathfrak{A},\mathfrak{B})=4\ln 2$ and
$I^{\left(s\right)}(\mathfrak{A},\mathfrak{B})=-4\ln 2$. In case
$\mu_1=\mu_2=\frac{\pi}{2}$, $\nu_1=\nu_2=0$ we have
$\theta_{11}=\theta_{12}=\theta_{21}=\theta_{22}=\frac{1}{4}$,
$e(\theta_{11})=e(\theta_{12})=e(\theta_{21})=e(\theta_{22})=0$,
and $I(\mathfrak{A},\mathfrak{B})=0$. Finally, in case
$\mu_1=\mu_2=\pi$, $\nu_1=\nu_2=0$ we have
$\theta_{11}=\theta_{12}=\theta_{21}=\theta_{22}=\frac{1}{2}$,
$e(\theta_{11})=e(\theta_{12})=e(\theta_{21})=e(\theta_{22})=1$,
and $I^{\left(s\right)}(\mathfrak{A},\mathfrak{B})=4\ln 2$.

Since the image of a compact and connected set via continuous
function $I(\mathfrak{A},\mathfrak{B})$ (respectively, the
continuous function
$I^{\left(s\right)}(\mathfrak{A},\mathfrak{B})$) is a compact and
connected subset of $\mathbb{R}$, we obtain that the range of
$I(\mathfrak{A},\mathfrak{B})$ (respectively,
$I^{\left(s\right)}(\mathfrak{A},\mathfrak{B})$) coincides with
the interval $[0,4\ln 2]$ (respectively,  with the interval
$[-4\ln 2,4\ln 2]$).

\subsection{Bell's Inequality}

\label{15.1}

The equality
$|\mathcal{A}_{\mu_1}\mathcal{B}_{\nu_1}+\mathcal{A}_{\mu_1}\mathcal{B}_{\nu_2}+
\mathcal{A}_{\mu_2}\mathcal{B}_{\nu_1}-\mathcal{A}_{\mu_2}\mathcal{B}_{\nu_2}|=
2$ yields (with an abuse of the probability theory) Bell's
inequality
\[
|\mathcal{E}(\mathcal{A}_{\mu_1}\mathcal{B}_{\nu_1})+\mathcal{E}(\mathcal{A}_{\mu_1}\mathcal{B}_{\nu_2})+
\mathcal{E}(\mathcal{A}_{\mu_2}\mathcal{B}_{\nu_1})-\mathcal{E}(\mathcal{A}_{\mu_2}\mathcal{B}_{\nu_2}|\leq
2,
\]
that is, $|b(\mu_1, \mu_2, \nu_1,\nu_2)|\leq 2$, where $b(\mu_1,
\mu_2,\nu_1,\nu_2)=\cos(\mu_1-\nu_1)+\cos(\mu_1-\nu_2)+\cos(\mu_2-\nu_1)-\cos(\mu_2-\nu_2)$.

J.~S.~Bell in~\cite{[5]} proves that if there exist "...additional
variables which restore to the (quantum) theory causality and
locality", then the above inequality is satisfied. Since
$I(\mathfrak{A},\mathfrak{B})=0$ is equivalent to the equalities
$|\mu_i-\nu_j|=\frac{\pi}{2}$, $i,j=1,2$, this yields $b=0$. Thus,
we obtain that if Bell's inequality is violated, then the total
information flow $I(\mathfrak{A},\mathfrak{B})$ is strictly
positive, that is, the experiments $\mathfrak{A}$ and
$\mathfrak{B}$  are informationally dependent.

\begin{examples}\label{15.1.1} {\rm Note that the results of all calculations
below are rounded up to the 7-th digit.

1) (Aspect's experiment)
$\mu_1=\frac{\pi}{8},\mu_2=\frac{3\pi}{8}$, $\nu_1=\frac{\pi}{4}$,
$\nu_2=0$. Then we obtain $\cos(\frac{\pi}{8})=0.9238795$,
$\cos(\frac{3\pi}{8})=0.3826834$, and therefore $b(\frac{\pi}{8},
\frac{3\pi}{8}, \frac{\pi}{4},0)= 2.3889551$. On the other hand,
$\theta_{11}=\theta_{12}=\theta_{21}=\frac{1}{2}\sin^2(\frac{\pi}{16})=0.0190301$,
$e(\theta_{11})=e(\theta_{12})=e(\theta_{21})=-0.0415353$,
$\theta_{22}=\frac{1}{2}\sin^2(\frac{3\pi}{16})=0.154329$,
$e(\theta_{22})=-0.1084492$. Hence we have
$I(\mathfrak{A},\mathfrak{B})=0.1615415$ and
$I^{\left(s\right)}(\mathfrak{A},\mathfrak{B})=-0.2330551$.

2) $\mu_1=\pi,\mu_2=\frac{2\pi}{3}$, $\nu_1=0$,
$\nu_2=\frac{\pi}{3}$. Then we have
$b(\pi,\frac{2\pi}{3},0,\frac{\pi}{3})=-2.5$. On the other hand,
$\theta_{11}=\frac{1}{2}\sin^2(\frac{\pi}{2})=0.5$,
$e(\theta_{11})=1$,
$\theta_{12}=\theta_{21}=\frac{1}{2}\sin^2(\frac{\pi}{3})=\frac{3}{8}=
0.375$, $e(\theta_{12})=e(\theta_{21})=0.1887219$,
$\theta_{22}=\frac{1}{2}\sin^2(\frac{\pi}{6})=\frac{1}{8}=0.125$,
$e(\theta_{22})=-0.1887219$. Hence we obtain
$I(\mathfrak{A},\mathfrak{B})=1.0855833$ and
$I^{\left(s\right)}(\mathfrak{A},\mathfrak{B})=1.1887219$

3) $\mu_1=\frac{\pi}{2},\mu_2=0$, $\nu_1=\frac{\pi}{4}$,
$\nu_2=\frac{3\pi}{4}$. Then $b(\frac{\pi}{2},0,\frac{\pi}{4},
\frac{3\pi}{4})=2\sqrt{2}$. On the other hand,
$\theta_{11}=\theta_{12}=\theta_{21}=\frac{1}{2}\sin^2(\frac{\pi}{8})=0.0732233$,
$e(\theta_{11})=e(\theta_{12})=e(\theta_{21})=-0.3994425$,
$\theta_{22}=\frac{1}{2}\sin^2(\frac{3\pi}{16})=0.154329$,
$e(\theta_{22})=-0.10844492$. Hence we have
$I(\mathfrak{A},\mathfrak{B})=0.9053727$ and
$I^{\left(s\right)}(\mathfrak{A},\mathfrak{B})=-0. 22827767$.

4) $\mu_1=\pi,\mu_2=0$, $\nu_1=0$, $\nu_2=\pi$. Then we have
$b(\pi,0,0,-\pi)=2$. On the other hand,
$\theta_{11}=\theta_{22}=\frac{1}{2}\sin^2(\frac{\pi}{2})=0.5$,
$e(\theta_{11})=e(\theta_{22})=1$,
$\theta_{12}=\theta_{21}=\frac{1}{2}\sin^2(0)=0$,
$e(\theta_{12})=e(\theta_{21})=-1$. Therefore we obtain
$I(\mathfrak{A},\mathfrak{B})=4\ln2=2.7725887=\max
I(\mathfrak{A},\mathfrak{B})$ and
$I^{\left(s\right)}(\mathfrak{A},\mathfrak{B})=0$.

5) $\mu_1=\frac{5\pi}{6},\mu_2=\frac{2\pi}{3}$,
$\nu_1=\frac{\pi}{3}$, $\nu_2=\frac{\pi}{2}$. In this case we have
$b(\frac{5\pi}{6},\frac{2\pi}{3},\frac{\pi}{3},
\frac{\pi}{2})=1-\frac{\sqrt{3}}{2}$. On the other hand,
$\theta_{11}=\frac{1}{2}\sin^2(\frac{\pi}{4})=\frac{1}{4}$,
$e(\theta_{11})=0$,
$\theta_{12}=\theta_{21}=\frac{1}{2}\sin^2(\frac{\pi}{6})=\frac{1}{8}$,
$e(\theta_{12})=e(\theta_{21})=-0.1887219$,
$\theta_{22}=\frac{1}{2}\sin^2(\frac{\pi}{12})=0.0334936$,
$e(\theta_{22})=-0.6453728$. Hence we have
$I(\mathfrak{A},\mathfrak{B})=0.7089624$ and
$I^{\left(s\right)}(\mathfrak{A},\mathfrak{B})=-0.267929$.

6) The link

\url{http://www.math.bas.bg/algebra/valentiniliev/}

\noindent contains a Java experimental implementation
"dependencemeasurements2" depending on five parameters: an
non-negative integer $n$ and four real numbers
$\mu_1,\mu_2,\nu_1,\nu_2$ from the closed interval $[0,\pi]$. One
can also find the description of this program at the above link.

 }

\end{examples}

Examples~\ref{15.1.1} and, especially, example 6), yield that the
relations between $I(\mathfrak{A},\mathfrak{B})$ and $b$, and
$I^{\left(s\right)}(\mathfrak{A},\mathfrak{B})$ and $b$ seem to be
random. Below we present an attempt to explain the uncertainty of
this relation by refereing to~\cite[5.4, 5.5]{[25]}. We define the
events
\[
U=\{(\mu_1,\mu_2,\nu_1,\nu_2)\in
[0,\pi]^4||b(\mu_1,\mu_2,\nu_1,\nu_2)|\leq 2\},
\]
\[
V=\{(\mu_1,\mu_2,\nu_1,\nu_2)\in
[0,\pi]^4|I(\mathfrak{A},\mathfrak{B})(\mu_1,\mu_2,\nu_1,\nu_2)\in
[0,2\ln 2]\},
\]
\[
VS=\{(\mu_1,\mu_2,\nu_1,\nu_2)\in
[0,\pi]^4|I^{\left(s\right)}(\mathfrak{A},\mathfrak{B})(\mu_1,\mu_2,\nu_1,\nu_2)\in
[0,4\ln 2]\},
\]
with complements $U^c$, $V^c$, and $VS^c$ in $[0,\pi]^4$. We
suppose that the probabilities $\alpha=\pr(U)$, $\beta=\pr(V)$,
and $\beta^{\left(s\right)}=\pr(VS)$ in the sample space
$[0,\pi]^4$ furnished with normalized Borel measure are known. The
probabilities $\tau=\pr(U\cap V)$ and
$\tau^{\left(s\right)}=\pr(U\cap VS)$ run through the closed
intervals
$I(\alpha,\beta)=[\max(0,\alpha+\beta-1),\min(\alpha,\beta)]$ and
$I(\alpha,\beta^{\left(s\right)})$, respectively. In case
$\tau=\min(\alpha,\beta)]$ (respectively,
$\tau^{\left(s\right)}=\min(\alpha,\beta^{\left(s\right)})]$)
there exists a relation of inclusion (up to a set of probability
$0$) between $U$ and $V$ (respectively, between $U$ and $VS$).
Otherwise, both conditional probabilities $\pr(V^c|U)$ and
$\pr(V|U^c)$ (respectively, $\pr(VS^c|U)$ and $\pr(VS|U^c)$) can
not be simultaneously as small as one wants (a kind of uncertainty
principle).

\begin{remark}\label{15.1.5} {\rm The
probabilities $\alpha=\pr(U)$, $\beta=\pr(V)$, $\tau=\pr(U\cap
V)$, and $\tau^{\left(s\right)}=\pr(U\cap VS)$ can be approximated
by using Examples~\ref{15.1.5}, 6). We draw a random sample $X$ of
size $n$ from the sample space $[0,\pi]^4$ and consider $X$ as a
sample space with $n$ equally likely outcomes. Then the
probabilities $\hat{\alpha}(n)$, $\hat{\beta}(n)$,
$\hat{\tau}(n)$, and $\widehat{\tau^{\left(s\right)}}(n)$ of the
traces of $U$,$V$, $U\cap V$, and $U\cap VS$ on $X$ (the sample
proportions) are unbiased estimators for $\alpha$, $\beta$,
$\tau$, and $\tau^{\left(s\right)}$ when $n$ is large.

Note that as an output of $n$ iterations of the random process
from example 6) we can also find: a) the approximation
$[L(n),J(n)]$ of the range of $I(\mathfrak{A},\mathfrak{B})$ and
the approximation $[LS(n),JS(n)]$ of the range of
$I^{\left(s\right)}(\mathfrak{A},\mathfrak{B})$ under the
condition $|b|\leq 2$, b) the above sample proportions, and c) the
approximations
$\pr(V^c|U)(n)=\frac{\hat{\alpha}(n)-\hat{\tau}(n)}{\hat{\alpha}(n)}$,
$\pr(V|U^c)(n)=\frac{\hat{\beta}(n)-\hat{\tau}(n)}{1-\hat{\alpha}(n)}$,
and
$\pr(VS^c|U)(n)=\frac{\hat{\alpha}(n)-\widehat{\tau^{\left(s\right)}}(n)}{\hat{\alpha}(n)}$,
$\pr(VS|U^c)(n)=\frac{\widehat{\beta^{\left(s\right)}}(n)-\widehat{\tau^{\left(s\right)}}(n)}{1-\hat{\alpha}(n)}$.

Below are the results obtained by drawing a random sample of size
$n=1000$:

$\hat{\alpha}(1000)=0.838$, $\hat{\beta}(1000)=0.704$,
$\hat{\tau}(1000)=0.614$,
$\widehat{\tau^{\left(s\right)}}(1000)=0.087$,
$\pr(V^c|U)(1000)=0.2673031$, $\pr(V|U^c)(1000)=0.5555555$,
$\pr(VS^c|U)(1000)=0.8961814$, $\pr(VS|U^c)(1000)=0.0185185$.

}

\end{remark}

\section*{Acknowledgements}

I thank Dimitar Guelev for making  an experimental Java
implementation of the evaluation of dependence of simultaneous
measurements and the approximation of various parameters via a
random process. The numerical examples thus produced were
invaluable for my work. I would like to express my gratitude to
administration of the Institute of Mathematics and Informatics at
the Bulgarian Academy of Sciences for creating perfect and safe
conditions of work.

\end{document}